\documentclass{iau}
\usepackage{graphicx}

\title[The magnetic field of Pollux] 
{Pollux: a stable weak dipolar magnetic field but no planet ?}

\author[Michel Auri\`ere, Renada Konstantinova-Antova, Olivier Espagnet \etal]   
{Michel Auri\`ere$^1$,
 Renada Konstantinova-Antova$^{2,1}$,
 Olivier Espagnet $^1$,
 Pascal Petit $^1$,
 Thierry Roudier $^1$,
 Corinne Charbonnel $^{3,1}$,
 Jean-Fran\c cois Donati $^1$
 \& Gregg A. Wade $^4$}

\affiliation{$^1$IRAP,Universit\'e de Toulouse \& CNRS, Toulouse, France \\ email: {\tt michel.auriere@irap.omp.eu} \\[\affilskip]
$^2$Institute of Astronomy and NAO, Bulgarian Academy of Sciences, Sofia, Bulgaria  \\[\affilskip]
$^3$Geneva Observatory, University of Geneva, Versoix, Switzerland \\ 
$^4$Department of Physics, Royal Military College of Canada, Kingston, Ontario, Canada } 

\pubyear{2014}
\volume{302}  
\pagerange{119--126}
\jname{Magnetic fields throughout stellar evolution}
\editors{M. Jardine, P. Petit\& H. Spruit, eds.}
\begin{document}

\maketitle

\begin{abstract}
Pollux is considered as an archetype of a giant star hosting a planet: its radial velocity (RV) presents sinusoidal variations with a period of about 590 d, which have been stable  for more than 25 years.
Using ESPaDOnS and Narval we have detected a weak (sub-gauss) magnetic field at the surface of Pollux and followed up its variations with Narval during 4.25 years, i.e. more than for two periods of the RV variations.  The longitudinal magnetic field is found to vary with a sinusoidal behaviour with a period close to that of the RV variations and with a small shift in phase. 
We then performed a Zeeman Doppler imaging (ZDI) investigation from the Stokes $V$ and Stokes $I$  least-squares deconvolution (LSD) profiles. A rotational period is determined, which is consistent with the period of variations of the RV. 
The magnetic topology is found to be mainly poloidal and this component almost purely dipolar. The mean strength of the surface magnetic field is about 0.7 G.
As an alternative to the scenario in which Pollux hosts a close-in exoplanet, we suggest that the magnetic dipole of Pollux can be associated with two temperature and macroturbulent velocity spots which could be sufficient to produce the RV variations. We finally investigate the scenarii of the origin of the magnetic field which could explain the observed properties of Pollux.
\end{abstract}

\firstsection 
\section{Introduction}
Pollux ($\beta$ Geminorum, HD 62509) is a well studied K0III giant neighbour of the sun. It is considered as an archetype of a giant star hosting a planet since its presents periodic sinusoidal radial velocity (RV) variations which have been stable during more than 25 years (\cite[Hatzes \etal\ 2006] {hatz06}). Now, \cite[Auri\`ere \etal\ (2009)] {aur09} discovered a weak magnetic field at the surface of Pollux whose variations could be correlated with the RV ones. We have therefore performed a Zeeman survey of Pollux and collected spectropolarimetric data during 4.25 years i.e. more than for two periods of the RV variations. 

\section{A spectropolarimetric survey of Pollux with Narval and ESPaDOnS}

From 2007 September to 2012 February, using first ESPaDOnS (\cite[Donati \etal\ 2006] {don06}) at CFHT in a snapshot program, then Narval (its twin) at the TBL in a systematic survey, we observed Pollux on 41 dates and got 266 Stokes $V$ series. 
 The observational properties of the instruments and reduction techniques are the same as described by \cite[Auri\`ere \etal\ (2009)] {aur09}. To obtain a high-precision diagnosis of the spectral line circular polarization, LSD (\cite[Donati \etal\ 1997] {don97}) was applied to each reduced Stokes $I$ and $V$ spectrum. 
From these mean Stokes profiles we computed the surface-averaged longitudinal magnetic field  $B_\ell$ in G, using the first-order moment method adapted to LSD profiles (\cite[Donati et al. 1997] {don97}). The RV of Pollux was measured from the averaged LSD Stokes $I$ using a gaussian fit. Figure 1 shows the variations of RV (left plot) and $B_\ell$ (right plot) during the 2007-2012 seasons of our survey:
The  $B_\ell$ variations appear to follow those of RV.

\begin{figure}[b]
\begin{center}
\includegraphics[width=2.6in]{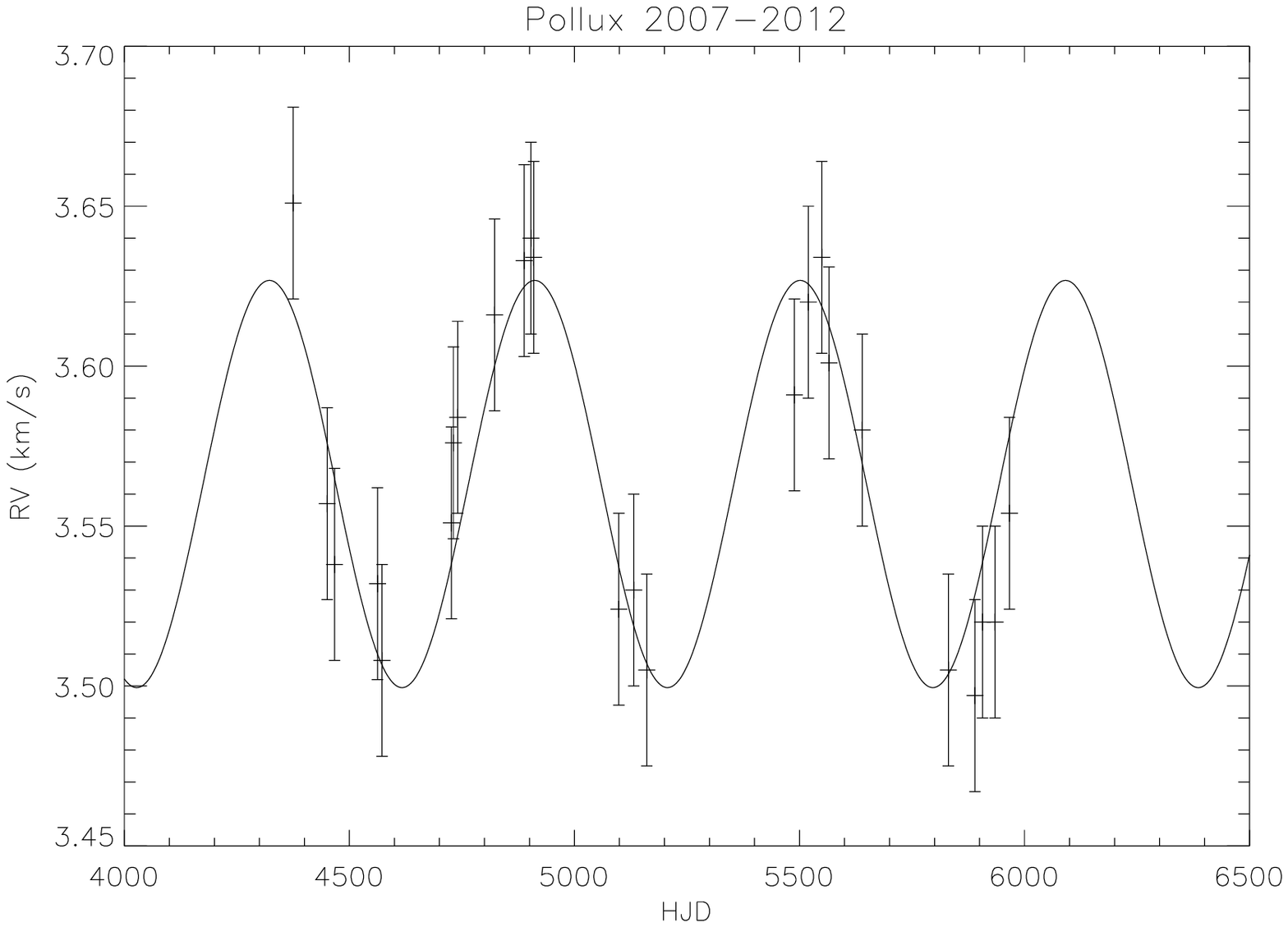}
\includegraphics[width=2.6in]{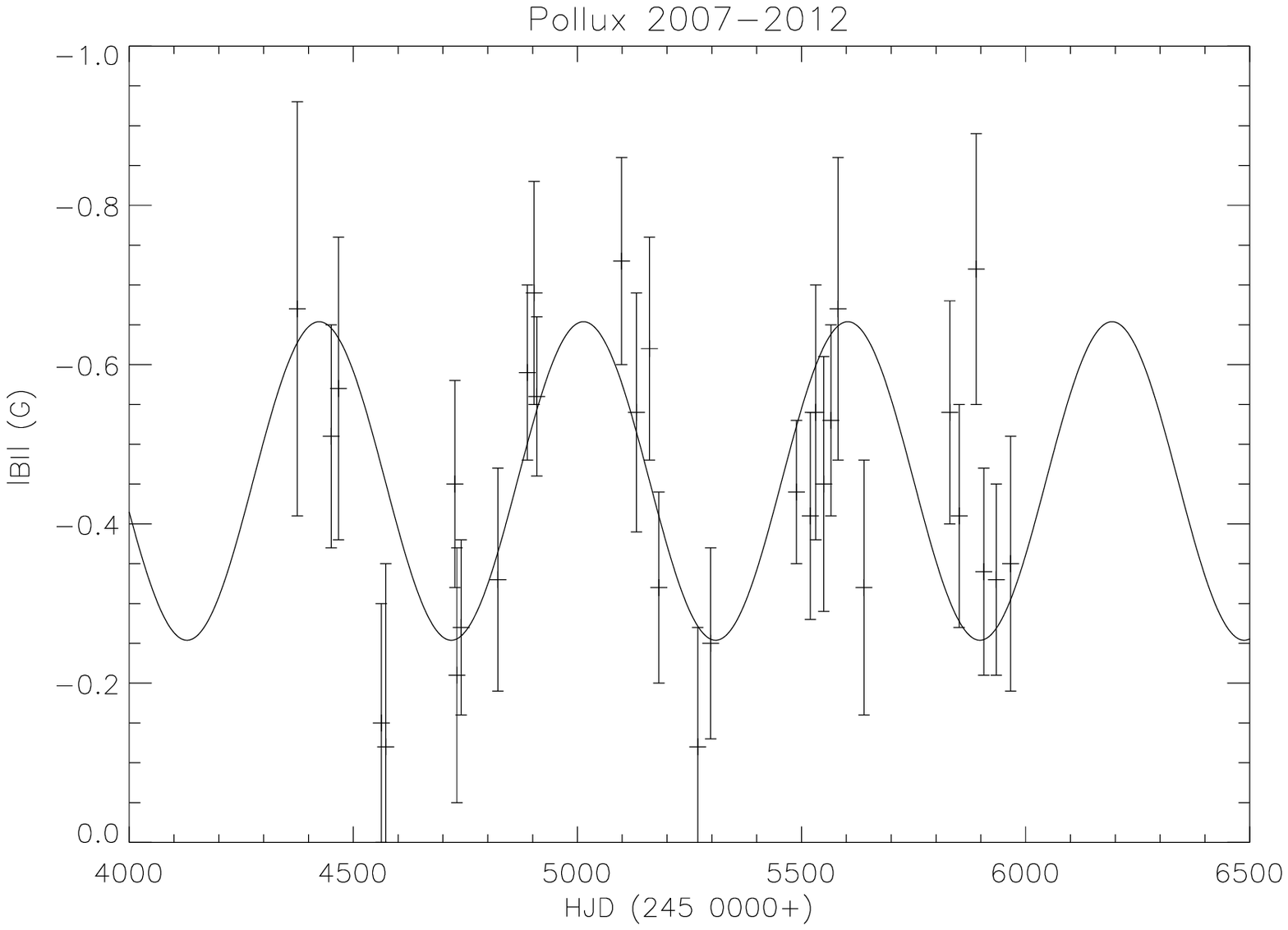}
 \caption{Variations of RV (left plot) and of $B_l$ (right plot) with HJD (245 0000+) in 2007-2012. A sinusoid with P= 589.64 d is fitted for each parameter.}
   \label{fig1}
\end{center}
\end{figure}

\section{Modeling the magnetic field of Pollux} 

{\underline{\it Variations of the longitudinal magnetic field }}.

Figure 1 shows the fit of the variations of RV and $B_\ell$ with a sinusoid of P = 589.64 d as derived by \cite[Hatzes \etal\ (2006)] {hatz06}. With this period and adjusting the amplitude of the $B_\ell$ variations to 0.2 G, the sinusoids obtained by least squares fitting have a phase shift of 146 d, i.e. about 25 \% of the period.

{\underline{\it Zeeman Doppler imaging (ZDI) of Pollux}}.

To use the whole Zeeman information included in our Stokes V  data, and to independently infer the rotational period $P_{rot}$ and the magnetic topology of Pollux,  we have used the ZDI method in the version of Donati et al. (2006). We limited the number of spherical harmonics to $l < 5$ since increasing the threshold did not change significantly the results. We have followed the approach of Petit et al (2002) for determining the $P_{rot}$ of Pollux, and we find a value of about 587 d as a favoured period. This is very near the RV variation period and both are possible $P_{rot}$ with respect to our error bars. Since the RV period has been found stable during more than 25 years, we consider that it is the real $P_{rot}$ and will use it hereafter in this work. With these parameters, our prefered ZDI model is the following: the inclination angle $i$ is 60$^\circ$ and the poloidal component contains 71 \%  of the reconstructed magnetic energy. The dipole component corresponds to about 99 \% of this poloidal magnetic energy. The mean magnetic field $B_{mean}$ is 0.7 G, and the angle between the rotation and magnetic dipole axis $\beta$ is about 20$^\circ$. In these conditions and within the solid rotation hypothesis, the $v\sin i $ of Pollux would be 0.7 km s$^{-1}$.

\begin{figure}[b]
\begin{center}
\includegraphics[width=2.35in]{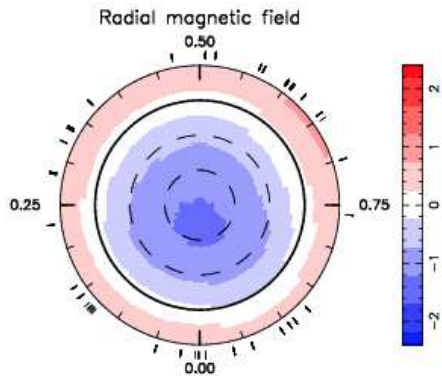}
 \caption{ZDI map of Pollux (radial component of the magnetic field) presented in flattened polar projection down to latitudes of -30$^\circ$. The magnetic field strength is expressed in G.}
   \label{fig1}
\end{center}
\end{figure}

\section {Origin of the radial velocity variations of Pollux}

Since we found that the $P_{rot}$ of Pollux is equal to the period of the RV variations, one has to consider if the weak magnetic field is able to induce these RV variations. 

{\underline{\it The planet hypothesis}}

If a planet exists, orbiting around Pollux and being responsible for the RV variations, the planet revolution is synchronized with the rotation of the star. From \cite[Hatzes \etal\ (2006)] {hatz06}, the mass of the planet is in the range 3-13 $M_{Jupiter}$ and the semi major axis is $a$ = 1.64 AU. In these conditions we are very far from the case occuring with ''hot jupiter'' class objects (e.g. \cite[Fares \etal\ 2013] {fares13}), and the expected tidal effects are too weak for enabling the synchronization (\cite[Pont 2009] {pont09}).

{\underline{\it The magnetic and macroturbulence spot hypothesis}}

 A dipole configuration as inferred for Pollux by our ZDI study can be regarded as corresponding to two  magnetic spots. 
  Using the SOAP online package (\cite[Boisse \etal\ 2012] {boisse12}) we have simulated the effects produced by the magnetic configuration of Pollux: we could generate the observed RV variations but photometric variations of larger amplitude than those detected in the Hipparcos data were predicted. Now, \cite[Hatzes \& Cochran (2000)] {hatz00} and \cite[Lee \etal\ (2012)] {lee12}, proposed the effect of macroturbulent velocity spots linked with a surface magnetic field  to explain the RV variations of Polaris and $\alpha$ Per respectively. RV variations up to 100 m s$^{-1}$ were expected for these stars without inducing photometric variations.
Consequently, we consider that the magnetic properties of Pollux can make the hosted-planet hypothesis unecessary.

\section {Origin of the magnetic field}

Pollux is the first giant with a sub-G mean surface magnetic field for which $P_{rot}$ is determined. This $P_{rot}$  is also the longest determined up to now for a giant.  If the hypothesis of a planet orbiting around Pollux is not retained, the very long stability of the dipolar magnetic field (more than 25 years or 15 $P_{rot}$) has to be taken into account to infer the origin of the magnetic field. The high Rossby number $Ro$ of 2-3 inferred for Pollux using this $P_{rot}$ and the convective turnover time $\tau_{conv}$ derived using the evolutionary models with rotation of \cite[Lagarde \etal\ (2012)] {lagarde12} and Charbonnel et al. (this symposium and in preparation) suggest that an $\alpha$-$\omega$ dynamo alone cannot be the origin of the observed magnetic field. Some other possibilities are:

- The high  $Ro$ and the stability would suggest Pollux being an Ap star descendant. However using $B_{mean}$ = 0.7 G and taking into account the magnetic flux conservation hypothesis we find that an Ap star progenitor of Pollux would have a magnetic strength of about 14 G, i.e. well below the 100-300 G lower limit found for Ap star dipole strength by \cite [Auri\`ere  \etal\ (2007)] {aur07}.

- An extreme case of dynamo when rotation is not involved is a local dynamo as inferred in the case of Betelgeuse (\cite [Auri\`ere  \etal\ 2010] {aur07}). However, in the case of Pollux the expected number of convective cells is much greater than inferred for Betelgeuse. Furthermore, in the case of Betelgeuse the magnetic field is observed to vary intrinsically on timescales of weeks to months.

- An intermediate case of dynamo that could operate during the red giant branch phase because of the deepening of the convective envelope, is the distributed dynamo. This type of dynamo is expected to occur in M dwarfs, and a long term stability has been reported in the case of the fast rotating and strongly magnetic star V374 Peg ( \cite [Morin  \etal\ 2008] {morin08}). One could suggest that a weak dynamo regime could be stable as well.

\section{Conclusion}

We have monitored the magnetic field of Pollux during 4.25 years. Our ZDI investigation shows that the $P_{rot}$ is equal to the RV variation period and that the Pollux surface magnetic field is mainly poloidal and dipolar. We then show that photometric and macroturbulent-velocity spots associated to the magnetic poles could explain the RV variations and could then make the hosted-planet hypothesis unnecessary. In this hypothesis, to explain the long term stability of the  RV variations, the magnetic field of Pollux should be stable during tens of years. We suggest that a weak-regime dynamo (of interface-type?) could be the origin of the magnetic field of Pollux.

In this way Pollux could be the representative of a class of weakly magnetic G K giants, recently discovered (Auri\`ere  \etal\ in preparation, Konstantinova-Antova \etal\, this symposium). These stars would include bright giants like Alphard and  Arcturus, and some of them, like Aldebaran and $\epsilon$ Tau also present stable  RV variations and are considered to host planets. Though we cannot exclude completely that neither Pollux hosts a planet in addition to its magnetic field, nor that it is a peculiar very stable magnetic star, our investigation suggests that Pollux could be the archetype of a  class of weakly magnetic G K giants.

\end{document}